\documentstyle[12pt]{article}
\begin{document}
\topmargin 0pt
\oddsidemargin 7mm
\headheight 0pt
\topskip 0mm
\addtolength{\baselineskip}{0.40\baselineskip}

\hfill June, 1995

\hfill EWHAGC 101/95

\hfill SNUTP 95-069

\hfill gr-qc/9506064

\begin{center}
\vspace{0.5cm}
{\large \bf Hawking Radiation and Evaporation of the Black Hole
Induced by a Klein-Gordon Soliton}
\end{center}
\vspace{1cm}

\begin{center}
Sung-Won Kim$^{(*)}$ and Won T. Kim$^{(\dagger)}$ \\
{\it Department of Science Education and Basic Science Research Institute,\\
Ewha Women's University, Seoul 120-750, South Korea}
\end{center}

\vspace{1cm}

\begin{center}
{\bf ABSTRACT}
\end{center}

A two-dimensional dilatonic black hole induced by a topological soliton
is exactly solvable in the scalar field theory coupled to dilaton gravity.
The Hawking radiation of the black hole is studied in the one-loop
approximation
with the help of the trace anomaly of energy-momentum
tensors which is a geometrical invariant.
The quantum theory can be also soluble in the RST scheme
in order to consider the back reaction of
the metric. The energy of the black hole system is calculated
and the classical thunderpop energy corresponding to the soliton
energy is needed to describe the final state of the black hole.
Finally we discuss the possibility of conservation of the topological
charge.

\vspace{2cm}
\hrule
\vspace{0.5cm}
\hspace{-0.6cm}$^{(*)}$ E-mail address : sungwon@ewhahp3.ewha.ac.kr \\
\hspace{-0.6cm}$^{(\dagger)}$ E-mail address : wtkim@ewhahp3.ewha.ac.kr \\
\vspace{1cm}

\newpage
\pagestyle{plain}

The dilaton gravity model coupled to
the conformal fields given by Callan-Giddings-Harvey-Strominger (CGHS) [1]
is very intriguing since this model possesses
most of the interesting properties of the
four-dimensional gravity theories such as a dynamical black hole solution and
Hawking radiations [2] even though the system has fewer degrees of freedom
compared to the four-dimensional gravity.
The infalling matter fields of a shock wave form a dynamical black hole,
and the back reaction of the metric due to the Hawking radiation
in the one-loop approximation is described by the conformal anomaly term
in the semiclassical action [1]. Russo-Susskind-Thorlacius (RST) [3] obtained
an exact dynamical black hole solution at the quantum level by adding
a local counter term to the CGHS model,
and the back reaction of the metric and the final state of black hole
have been well appreciated.

Recently, a two-dimensional dilaton gravity coupled to the solitons [4]
was proposed as a generalization of the infalling shock wave
and exactly solvable models.
The massive infalling soliton induces the classical black hole solution.
The quantum aspects of the black hole induced by the soliton
especially for the case of
Sine-Gordon model were studied by considering the Hawking
radiation on the black-hole background in Ref. [5].

In this paper, we introduce a time-dependent classical black hole solution
induced by the Klein-Gordon soliton having a lumped energy,
which satisfies asymptotic flatness in our spacetime.
Then the quantum-mechanical black hole solution incorporated with
the back reaction of the metric due to the Hawking radiation is
semiclassically obtained in the one-loop approximation.
To obtain the exact solution, we use the fact that the trace anomaly of
the matter fields is geometrically invariant quantity and
the effective action may have a local counter term.
The Hawking radiation and the Bondi energy reflecting the back reaction
of the metric are calculated in the future null infinities
and the final state of the black hole is discussed.
Finally we debate the conservation of the topological soliton charge.

Let us now introduce the two-dimensional dilaton gravity coupled to
the matter field,
\begin{eqnarray}
S_{\rm Cl} &=& S_{\rm DG} + S_{\rm f} \label{caction}, \\
S_{\rm DG}&=& \frac{1}{2\pi} \int d^2 x \sqrt{-g} \left[e^{-2\phi} (R + 4 ({\bf
                \bigtriangledown} \phi)^2 +4\lambda^2) \right], \nonumber \\
S_{\rm f}&=& \frac{1}{2\pi} \int d^2 x \sqrt{-g} \left[ -\frac{1}{2}
                 (\bigtriangledown f)^2  -e^{-2\phi}U(f) \right]  \nonumber
\end{eqnarray}
where $g$, $\phi$, and $f$ are metric, dilaton, and matter fields
respectively.
The classical equations of motion are given by
\begin{eqnarray}
e^{-2\phi} \hspace{-0.5cm}&&\hspace{-0.5cm}
\left[ R_{\mu \nu} + 2 \bigtriangledown_\mu \bigtriangledown_\nu
\phi \right] = \frac{1}{2} \left( \bigtriangledown_\mu f \bigtriangledown_\nu
f - \frac{1}{2} g_{\mu \nu} ( \bigtriangledown f)^2 \right), \\
R \hspace{-0.2cm}&=&\hspace{-0.2cm} 4 \left[ (\bigtriangledown \phi)^2 -
\bigtriangledown^2 \phi +\lambda^2 \right] + U(f), \\
\bigtriangledown^2 f \hspace{-0.3cm}&-&\hspace{-0.3cm} e^{-2\phi} \frac{\delta
U(f)}{\delta f} =0.
\end{eqnarray}
In the conformal gauge, $g_{\pm \mp}=-\frac{1}{2} e^{2\rho}$ and
$g_{\pm \pm}=0$, the equations of motion then reduce to
\begin{eqnarray}
T_{+-} &=& 2 e^{-2\phi} \partial_+ \partial_- ( \rho -\phi)=0, \label{+-}\\
T_{\pm \pm} &=& e^{-2\phi} \left( -2 \partial_{\pm}^2 \phi + 4 \partial_\pm
\rho
\partial_\pm \phi \right) + \frac{1}{2} (\partial_\pm f)^2 =0, \label{two}\\
-4 \partial_+ \partial_- \phi \hspace{-0.2cm}&+&\hspace{-0.2cm} 4 \partial_+
\phi \partial_- \phi +2 \partial_+
\partial_- \rho +\left( \lambda^2  -\frac{1}{4} U(f) \right) e^{2\rho}=0,\\
\partial_+ \partial_- f \hspace{-0.2cm}&+&\hspace{-0.2cm}\frac{1}{4} e^{ 2(\rho
-\phi)} \frac{\delta U(f)}
{\delta f} =0.
\end{eqnarray}
In Eq. (\ref{+-}), we obtain the solution,
\begin{equation}
\rho=\phi +\frac{1}{2} \left( \omega_+(x^+) + \omega_- (x^-) \right)
\end{equation}
where the holomorphic and antiholomorphic functions $\omega_{\pm}$ reflect
the residual conformal degrees of freedom.
In the Kruskal gauge $\omega_{\pm}=0$,
the equations of motion are most simply given by
\begin{eqnarray}
\partial_{\pm}^2 (e^{-2 \phi}) \hspace{-0.2cm}&+&\hspace{-0.2cm}
\frac{1}{2}(\partial_\pm f)^2 =0,
\label{pmpm} \\
\partial_+ \partial_- (e^{-2\phi})\hspace{-0.2cm} &+&\hspace{-0.2cm} \lambda^2
- \frac{1}{4} U(f) =0,
\label{dilatoneq}\\
\partial_+ \partial_- f\hspace{-0.2cm}&+&\hspace{-0.2cm} \frac{1}{4}
\frac{\delta U(f)}{\delta f} =0.
\label{solitoneq}
\end{eqnarray}

Considering the potential
$U(f)=\frac{\beta}{4} (f^2 -\frac{\mu^2}{\beta} )^2$
in order to obtain the Klein-Gordon soliton solution [4]
where the coupling constants $\mu$ and $\beta$ are positive,
the Klein-Gordon equation (\ref{solitoneq}) yields an exact solution,
\begin{equation}
\label{fsol}
f(x^+, x^-) = \sqrt{\frac{\mu^2}{\beta}}  {\rm tanh}( \Delta -\Delta_0),
\end{equation}
where $\Delta(x^+,x^-)= \frac{\mu}{2\sqrt{2}} (\alpha x^+ - \frac{1}{\alpha}
x^-)$, $\alpha=\sqrt{\frac{1+v}{1-v}}$, $-1<v<1$,
and $x^\pm_0$ implies
the center of soliton coordinate. The range of infalling
velocity of the soliton is $0<v_{\rm in}< 1$ and outgoing case is
$-1<v_{\rm out}<0$. Accordingly we denote $\Delta_{\rm in}$
or $\Delta_{\rm out}$ depending on the direction of velocity of soliton.
{}From Eqs. (\ref{pmpm}) and (\ref{dilatoneq}), the infalling soliton
soution (\ref{fsol})
induces dilaton and metric solution, which is given by
\begin{eqnarray}
e^{-2\phi (x^+, x^-)}& =& e^{-2\rho (x^+, x^-)} \nonumber\\
&=&C + a_+x^+ +a_-x^- -\lambda^2 x^+ x^-   \nonumber \\
&&-\frac{\mu^2}{12\beta} {\rm tanh}^2 (\Delta_{\rm in} -\Delta_0)
-\frac{\mu^2}{3\beta} {\rm lncosh}(\Delta_{\rm in} -\Delta_0)
\end{eqnarray}
where $C$ and $a_{\pm}$ are constants to be determined by
boundary conditions.

For the limit $\Delta_{\rm in} -\Delta_0 \ll -1$, we require the metric-dilaton
solution to be an exact linear dilaton vacuum (LDV).
By choosing the above constants as
\begin{equation}
C=\frac{\mu^2}{3\beta}( \Delta_0 +\frac{1}{4} -\ln 2),~~
a_+=-\frac{\mu^3\alpha_{\rm in}}{6\sqrt{2} \beta},~~
a_-=+\frac{\mu^3}{6\sqrt{2} \alpha_{\rm in} \beta},
\end{equation}
we obtain the LDV
\begin{equation}
e^{-2\phi} = -\lambda^2 x^+ x^- .
\end{equation}
For the limit $\Delta_{\rm in} -
\Delta_0 \gg 1$, the definite black hole geometry appears
as
\begin{equation}
e^{-2\phi} = \frac{M}{\lambda} -\lambda^2
\left( x^+ - \frac{\mu^3}{3\sqrt{2} \alpha_{\rm in} \beta \lambda^2} \right)
\left( x^- + \frac{\mu^3 \alpha_{\rm in}}{3\sqrt{2} \beta \lambda^2} \right)
\end{equation}
where $M=\frac{\mu^2 \lambda}{3\beta} \left( 2\Delta_0 -\frac{\mu^4}{6\beta
\lambda^2} \right)$ is a ADM mass [6,17].
We assume that the black hole mass $M$ is positive definite {\it i.e.},
\begin{equation}
\alpha_{\rm in} x^+_0 - \frac{1}{\alpha_{\rm in}} x^-_0 >
\frac{\mu^3}{3\sqrt{2} \beta
\lambda^2}
\end{equation}
which is more explicitly, depending on the velocity of the infalling soliton,
for $v_{\rm in} \rightarrow 0$, $x_0^1 > \frac{\mu^3}{3\sqrt{2}\alpha_{\rm in}
\beta \lambda^2} > 0 $
and for $v_{\rm in} \rightarrow 1$, then $x_0^+ >
\frac{\mu^3}{3\sqrt{2}\alpha_{\rm in}\beta\lambda^2} >0 $.
It means that the center of soliton $x^\pm_0$ can be located
in our spacetime.

It seems to be appropriate to comment on the two limits.
For $\Delta_{\rm in} -\Delta_0 \ll-1 $  equivalently
$\alpha_{\rm in} x^+ - \frac{1}{\alpha_{\rm in}} x^-  \ll
\alpha_{\rm in} x^+_0 - \frac{1}{\alpha_{\rm in}} x^-_0 $,
this means that the region we consider is restricted to
$x^1 \ll  x^1_0$ for the small
velocity of soliton or $x^+ \ll  x^+_0$ for the speed of light.
For $\Delta_{\rm in} -\Delta_0 \gg 1$, the region is located in
$x^1 \gg x^1_0$ for $v_{\rm in}
\rightarrow 0$ and $x^+ \gg x^+_0$ for $v_{\rm in} \rightarrow 1$.

To study quantum effects of black hole,
we follow  Wald's axioms [5,7].
The two-dimensional trace of energy-momentum
tensors is determined only by the curvature scalar $R$ which is
the only available geometric invariant,
\begin{equation}
\label{anomaly}
T^\mu_\mu= \frac{1}{2} \kappa R
\end{equation}
where $\kappa$ is a constant which may be determined by some explicit
calculation of trace anomaly.
On the other hand, the ghost contribution to the trace anomaly (\ref{anomaly})
in the conformal gauge fixing can be canceled
by adding a ghost decoupling term [8,9,14],
$S_{St}=\frac{1}{\pi}\int d^2x \left[2\partial_+(\rho -\phi)
\partial_-(\rho -\phi) \right]$.
Anyway we assume that $\kappa$ is positive to obtain
a positive definite Hawking
radiation.

We now consider the back reaction of the metric semiclassically
by including the
effect of trace anomaly (\ref{anomaly}) to the classical action.
The quantum effective action is assumed to be
\begin{equation}
\label{qteff}
S_{\rm Qt}= \frac{1}{2\pi} \int d^2 x \sqrt{-g} \left[
-\frac{\kappa}{4} R \frac{1}{\bigtriangledown^2} R -\frac{\gamma}{2} \phi R
\right]
\end{equation}
where we include a local counter term with a constant $\gamma$.
In the conformal gauge, the total action reduces to
\begin{eqnarray}
\label{total}
S_{\rm T} &=& S_{\rm Cl} + S_{\rm Qt}     \nonumber \\
  &=&\frac{1}{\pi} \int d^2 x \left[
     e^{-2\phi} \left( 2\partial_+ \partial_-
\rho -  4\partial_+ \phi \partial_- \phi +  (\lambda^2 -\frac{1}{4}U(f))
e^{2\rho} \right)
\right. \nonumber  \\
&+& \left. \frac{1}{2}
\partial_+ f \partial_- f
-\kappa \partial_+ \rho \partial_-
\rho   -\kappa \phi \partial_+ \partial_-
\rho \right]
\end{eqnarray}
where we choose $\gamma =\kappa$ to solve exactly.
Following the Bilal and Callan [10] and de Alwis's method [11], we perform
field redefinition to a Liouville-like theory [3],
\begin{eqnarray}
\Omega &=&\frac{\sqrt{\kappa}}{2} \phi
   +\frac{e^{-2\phi}}{\sqrt{\kappa}},  \nonumber    \\
\chi &=& \sqrt{\kappa } \rho -\frac{\sqrt{\kappa}}{2} \phi +
\frac{e^{-2\phi}}{\sqrt{\kappa}}.
\end{eqnarray}
Then, the action (\ref{total}) and the two constraint
equations (\ref{two}) in terms of the redefined fields
are given by
\begin{eqnarray}
S_{\rm T}\hspace{-0.3cm} &=&\hspace{-0.3cm}
\frac{1}{\pi} \int d^2 x  \left[ -\partial_+ \chi \partial_- \chi
+ \partial_+ \Omega \partial_- \Omega
+ (\lambda^2 -\frac{1}{4}U(f)) e^{\frac{2}{\sqrt{\kappa}}(\chi -\Omega)}
+ \frac{1}{2}  \partial_+ f \partial_- f \right],
\label{newtotal}\\
\kappa t_\pm \hspace{-0.3cm}  &=&\hspace{-0.3cm}
-\partial_\pm \chi \partial_\pm \chi + \partial_\pm \Omega
\partial_\pm \Omega
+\sqrt{\kappa} \partial^2_\pm \chi + \frac{1}{2}
\partial_\pm f \partial_\pm f, \label{newpmpm}
\end{eqnarray}
where $t_{\pm}$ reflect the nonlocality of the quantum effective action
(\ref{qteff}).
{}From the action (\ref{newtotal}), we obtain  equations of motion,
\begin{eqnarray}
\partial_+ \partial_-\chi + \frac{1}{\sqrt{\kappa}}
(\lambda^2 -\frac{1}{4} U(f) )
e^{\frac{2}{\sqrt{\kappa}} (\chi -\Omega)}&=& 0,    \\
\partial_+ \partial_-\Omega + \frac{1}{\sqrt{\kappa}}
(\lambda^2 -\frac{1}{4} U(f) )
e^{\frac{2}{\sqrt{\kappa}} (\chi -\Omega)}&=& 0,     \\
\partial_+ \partial_- f +\frac{1}{4} e^{\frac{2}{\sqrt{\kappa}}
(\chi -\Omega)} \frac{\delta U(f)}{\delta f} &=&0.
\end{eqnarray}
In the Kruskal gauge, we obtain
the following solution,
\begin{eqnarray}
\label{omegachi}
\Omega(x^+,x^-)&=& \chi(x^+,x^-)  \nonumber \\
               &=& \frac{1}{\sqrt{\kappa}}
\left[ C + a_+x^+ + a_- x^- -\lambda^2 x^+ x^- -\frac{\kappa}{4}\ln(-\lambda^2
x^+
x^-) \right.    \nonumber \\
&&\left. -\frac{\mu^2}{12\beta} \tanh^2 (\Delta_{\rm in} -\Delta_0)
-\frac{\mu^2}{3\beta} \rm{ln cosh}(\Delta_{\rm in} -\Delta_0)\right]
\end{eqnarray}
where the solution obeys the constraints (\ref{newpmpm})
if $t_\pm =0$.

To delineate the geometry given by (\ref{omegachi}), we consider
asymptotic regions, in fact, the physically relevant
quantities such as Hawking radiation and Bond energy [12]
can be defined in these regions.
In the asymptotic region, $\Delta_{\rm in} -\Delta_0 \ll -1$,
the solution $\Omega$ becomes
the LDV,
\begin{equation}
\label{ldv}
\bar{\Omega}= -\frac{\lambda^2}{\sqrt{\kappa}} x^+ x^- -\frac{\sqrt{\kappa}}
{4} \ln(-\lambda^2 x^+ x^-).
\end{equation}
On the other hand, for $\Delta_{\rm in} -\Delta_0 \gg1$,
the definite black-hole geometry appears as
\begin{equation}
\label{omega}
\Omega= \frac{M}{\sqrt{\kappa}\lambda}-\frac{\sqrt{\kappa}}{4} \ln(-\lambda^2
x^+x^-)
-\frac{\lambda^2}{\sqrt{\kappa}} \left(x^+ -\frac{\mu^3}{3\sqrt{2}\alpha_{\rm
in}\beta
\lambda^2}\right) \left(x^- +\frac{\mu^3 \alpha_{\rm
in}}{3\sqrt{2}\beta\lambda^2}
\right),
\end{equation}
where in the present limit, the region we consider is
$x^1 \gg \frac{\mu^3}{3\sqrt{2}\alpha_{\rm in}
\beta \lambda^2}$
or $x^+ \gg \frac{\mu^3}{3\sqrt{2}\alpha_{\rm in}\beta\lambda^2}$.

The singularity occurs at the boundary of the range of $\Omega$
where $\Omega_{\rm min} =\frac{\sqrt{\kappa}}{4}(1-\ln\frac{\kappa}{4})$,
which is given by
\begin{eqnarray}
\frac{\kappa}{4}(1-\ln\frac{\kappa}{4})&\hspace{-0.4cm} =
\hspace{-0.4cm}&-\lambda^2 \bar{x}^+(\bar{x}^- +\frac{\mu^3\alpha_{\rm
in}}{3\sqrt{2}\beta\lambda^2})
-\frac{\kappa}{4}\ln(-\lambda^2
\bar{x}^+ \bar{x}^-)    \nonumber \\
&+&\frac{\mu^3}{3\sqrt{2}\alpha_{\rm in}\beta}\bar{x}^-
+\frac{2\mu^2 \Delta_0}{3\beta}.                  \label{singularity}
\end{eqnarray}
The location of the singularity is inside
an apparent horizon which is given by another curve
$\partial_+
\Omega =0$,
\begin{equation}
\label{apparent}
\lambda^2 \hat{x}^+(\hat{x}^- +\frac{\mu^3 \alpha_{\rm in}}{3\sqrt{2}\beta
\lambda^2})
+\frac{\kappa}{4}=0.
\end{equation}
Following the suggestion of Hawking [13], RST showed that the singularity
and apparent horizon collide in a finite proper
time and the singularity
is naked after the two have merged [3].
{}From (\ref{singularity}) and (\ref{apparent}),
the following relation for the intersection point $(x^+_s,x^-_s)$
is given
\begin{equation}
\frac{\kappa}{4} + \frac{\mu^3\alpha_{\rm in}}{3\sqrt{2}\beta} x^+_s =
\frac{\kappa}{4} e^{\frac{4M}{\kappa\lambda}} e^{-\frac{\mu^3\kappa}
{3\sqrt{2} \alpha_{\rm in}\beta\lambda^2 x^+_s}}
\end{equation}
and $x^-_s$ coordinate is determined by Eq. (\ref{apparent}).

In eq. (\ref{omega}), the first parenthesis of the third term
is approximately written by
\begin{equation}
x^+\left( 1 -\frac{\mu^3}{3\sqrt{2}\alpha_{\rm in}\beta\lambda^2 x^+} \right)
\approx x^+
\end{equation}
if we mainly treat the high speed limit of the infalling soliton
for convenience. Hereby,
the geometry can be very similar to that of the RST model.
We shall from now on consider the high speed case of
$\Delta_{\rm in} -\Delta_0 \gg 1$.
Since we now consider the asymtotic region $\Delta_{\rm in} -\Delta_0 \gg 1$
which is compatible with the range
$x^+ \gg x^+_0 > \frac{\mu^3}{3\sqrt{2} \alpha_{\rm in}\beta\lambda^2}$,
the intersection point is approximately given by
\begin{eqnarray}
\label{intersection}
x^+_s &\approx& \frac{3\sqrt{2}\beta\kappa}{4\mu^3\alpha_{\rm in}}
( e^{\frac{4M}{\kappa \lambda}} -1),
\nonumber \\
x^-_s &\approx& -\frac{\mu^3\alpha_{\rm in}}{3\sqrt{2}\beta\lambda^2}
\frac{1}{(1-e^{-\frac{4M}{\kappa
\lambda}})}.
\end{eqnarray}
Note that Eq. (\ref{intersection}) is valid only for
\begin{equation}
\label{mass}
M \gg \frac{\kappa \lambda}{4} \ln \left( 1+ \frac{2\mu^6}{9\beta^2
\lambda^2\kappa}
\right)
\end{equation}
since $x^+_s$ belongs to the region $x^+ \gg \frac{\mu^3}{3\sqrt{2}
\alpha_{\rm in}\beta\lambda^2}$.

For the finial state of the black
hole, a vacuum as a boundary condition
is chosen in such a way that the continuity condition is satisfied
between a shifted vacuum solution and the curved spacetime along some curve,
where a shifted vacuum is chosen as
\begin{eqnarray}
\bar{\Omega}&=& -\frac{\lambda^2}{\sqrt{\kappa}}
(x^+ -\frac{\mu^3}{3\sqrt{2}
\alpha_{\rm in}\beta\lambda^2})(x^- + \frac{\alpha_{\rm
in}\mu^3}{3\sqrt{2}\beta\lambda^2}) \nonumber \\
&&-\frac{\sqrt{\kappa}}{4} \ln\left(-\lambda^2
(x^+ -\frac{\mu^3}{3\sqrt{2} \alpha_{\rm in}\beta\lambda^2} )
(x^- +\frac{\alpha_{\rm in}\mu^3}{3\sqrt{2}\beta\lambda^2} )\right).
\end{eqnarray}
Along with the following curve, the geometry is continuous,
\begin{equation}
\frac{x^+x^-}{(x^+ -\frac{\mu^3}{3\sqrt{2} \alpha_{\rm in}\beta\lambda^2} )
(x^- +\frac{\alpha_{\rm in}\mu^3}{3\sqrt{2}\beta\lambda^2} )}
=e^{\frac{4M}{\kappa\lambda}}
\end{equation}
where
for $x^+ \gg \frac{\mu^3}{3\sqrt{2}\alpha_{\rm in}\beta\lambda^2}$,
the line is approximately straight line, $x^- \approx x^-_s$.
Then the energy density of thunderpop is
\begin{eqnarray}
\label{thunderpop}
\frac{1}{2} (\partial_- f)^2 &=& \kappa t_- -\sqrt{\kappa} \partial_-^2
\chi\nonumber \\
&\approx& -\frac{\kappa}{4} \frac{ (1-e^{-\frac{4M}{\kappa\lambda}})}{(x^-
+\frac{\mu^3\alpha_{\rm in}}{3\sqrt{2}\beta\lambda^2})}
                             \delta(x^- -x^-_s).
\end{eqnarray}

Before working out the Hawking radiation,
we discuss coordinate transformations.
The conformal transformations defined by
$x^{\pm} = \pm \frac{1}{\lambda} e^{\pm \lambda \sigma^\pm}$
do not give an asymptotically static configuration
and in particular the
dilaton and graviton fields do not approach
the correct form of LDV at infinity,
so we introduce
a quasi-static coordinate $y^\pm$ where the fields approach LDV in spatial and
null infinities,
\begin{eqnarray}
\label{static}
x^+ &=& \frac{1}{\lambda} e^{\lambda y^+} +
\frac{\mu^3}{3\sqrt{2} \alpha_{\rm in}
\beta \lambda^2}, \nonumber \\
x^- &=& -\frac{1}{\lambda} e^{-\lambda y^-} -\frac{\mu^3\alpha_{\rm
in}}{3\sqrt{2}
\beta \lambda^2}.
\end{eqnarray}
In this coordinate, $\Omega$ and $\chi$ are static to the leading order.
We shall calculate the Hawking radiation and Bondi energy in the
asymptotically static coordinate.

Let us now consider the Hawking radiation. From
the fundamental condition that $T_{\pm\pm}$ must be true tensors
without anomaly, we require the anomalous transformation as
\begin{equation}
t_\pm (y^\pm) = \left( \frac{\partial y^\pm}{\partial \sigma^\pm}\right)^{-2}
\left( t_\pm (\sigma^\pm) - \frac{1}{2} D^s_{\sigma^\pm} (y^\pm) \right)
\end{equation}
where the Schwarzian derivative is
$D^s_{\sigma^\pm}(y^\pm) = \frac{y^{-'''}}{y^{-'}}-
\frac{3}{2} (\frac{y^{-''}}{y^{-'}})^2$.
Then, following [10], we obtain the Hawking radiation,
\begin{eqnarray}
h(y^-) &=&-\kappa t_-(y^-)  \nonumber \\
&=&\frac{\kappa \lambda^2}{4}
\left[ 1- \frac{1}{(1+\frac{\mu^3\alpha_{\rm in}}{3\sqrt{2}\beta\lambda}
e^{\lambda y^- })^2} \right]
\label{h}
\end{eqnarray}
for $y^- < y^-_s$, and $h(y^-) =0$ for $y^- > y^-_s$.
As expected, for $y^- \to -\infty$, there is no Hawking radiation.
For $y^- < y^-_s$, the integrated Hawking flux $H(y^-)$ is calculated
as
\begin{eqnarray}
H(y^-)&=&\int^{y^-}_{-\infty} dy^- h(y^-) \nonumber \\
      &=&\frac{\kappa \lambda}{4}\left[ 1-\frac{1}{ (1+\frac{\mu^3\alpha_{\rm
in}}
      {3\sqrt{2}\beta\lambda}
      e^{\lambda y^- }) } +\ln(1+\frac{\mu^3\alpha_{\rm
in}}{3\sqrt{2}\beta\lambda}
      e^{\lambda y^- })  \right] \label{hr}.
\end{eqnarray}
For the limit, $y^- \rightarrow y^-_s-0$, we obtain
\begin{equation}
\label{hrs}
H(y^-_s-0) \approx M +\frac{\kappa \lambda}{4}(1 -e^{-\frac{4M}{\kappa \lambda}
} )
\end{equation}
which is greater than the total mass of the black hole.
Therefore, the remaining mass of black hole at this point is negative.
The integrated Hawking flux $H(y^-)$ is saturated after the
black hole is completely evaporated and
the total Hawking flux is a just $H(y^-_s).$
We approximately evaluated the integrated Hawking flux since
we did not obtain the exact intersection point.

It is convenient to write the Bondi energy as [14]
\begin{equation}
B(y^-) = \sqrt{\kappa} (\lambda +\partial_- -\partial_+)\delta \Omega
(y^+,y^-)|_{y^+ \rightarrow +\infty}       \label{bondi}
\end{equation}
where $\delta\Omega(y^+,y^-)=\Omega(y^+,y^-)-\bar{\Omega}(y^+,y^-)$.
Note that the contribution to the Bondi energy from the past
null infinity is zero since for $x^+ \rightarrow -\infty$,
the geometry is LDV as we see from (\ref{ldv}).
By putting Eq. (\ref{omega}) into (\ref{bondi}) in the asymptotically
static coordinates (\ref{static}),
we obtain
Bondi energy,
\begin{equation}
\label{bondi(y)}
B(y^-)= M -\frac{\kappa \lambda}{4} \left[ \frac{ \frac{\mu^3\alpha_{\rm
in}}{3\sqrt{2}\beta\lambda}}{(
\frac{\mu^3\alpha_{\rm in}}{3\sqrt{2}\beta\lambda} + e^{-\lambda y^- })}
+\ln(1 +\frac{\mu^3\alpha_{\rm in}}{3\sqrt{2}\beta\lambda} e^{\lambda y^-}
)\right].
\end{equation}
Note that at the point $y^-_s-0$, the Bondi energy is approximately given by
\begin{equation}
\label{bondi(s)}
B(y^-_s-0) \approx -\frac{\kappa \lambda}{4}(1-e^{-\frac{4M}{\kappa \lambda}})
\end{equation}
which is negative [15].
For $y^- >y^-_s$, the Bondi energy is zero since the negative energy of
black hole is emitted through the thunderpop (\ref{thunderpop})
and the final state of
black hole becomes the shifted LDV.
On the other hand, from Eqs. (\ref{hr}) and (\ref{bondi(y)}), we see that
the sum of integrated Hawking radiation and Bondi energy is conserved,
and independent of time. Naturally we see that the total
energy (ADM) [17] is conserved through the evaporation of the black hole.

In our black hole system, there are two kinds of
outgoing radiations, classical thunderpop and quantum-mechanical
Hawking radiation.
Further, we see that the carrier of classically infalling energy
is the infalling soliton
and may naturally expect that the outgoing classical thunderpop
is an outgoing soliton.
Let us now check briefly whether or not the thunderpop energy can be
the soliton energy.
The outgoing soliton of negative energy is given by
\begin{equation}
T^f_{--}\equiv -\frac{1}{2} (\partial_- f)^2
=-\frac{\mu^4}{16\beta\alpha^2_{\rm out}}
{\rm sech}^4 (\Delta_{\rm out} -\Delta_s)
\end{equation}
where $\Delta_s=\Delta( x^+_s,x^-_s)$.
The energy density falls off zero at the asymptotically null infinity.
In the shock wave limit, however, it reaches when
$v_{\rm out}$ approaches the velocity
of light where the soliton energy density is given by
\begin{equation}
T_{--}^f (x^+,x^-) \rightarrow -\frac{\mu^3}{3\sqrt{2}\beta\alpha_{\rm out}}
\delta (x^- -x^-_s).
\end{equation}
In terms of the asymptotically static coordinates (\ref{static}),
the soliton energy is
given by
\begin{equation}
\int dy^- T_{--}^f (y^+,y^-) = -\frac{\mu^3}{3\sqrt{2}\beta\alpha_{\rm out}}
e^{-\lambda y^-_s}.
\end{equation}
In our model, the soliton energy can be identified with
the thunderpop energy if the following condition is met,
\begin{equation}
-\frac{\mu^3}{3\sqrt{2}\beta\alpha_{\rm out}}
e^{-\lambda y^-_s} = -\frac{\kappa\lambda}{4}(1- e^{-\frac{4M}{\kappa\lambda}})
\end{equation}
and it reduces to
\begin{equation}
{\rm cosh}(\frac{4M}{\kappa\lambda})= 1 + \frac{\mu^6}{9\beta^2\lambda^2\kappa}
\left( \frac{\alpha_{\rm in}}{\alpha_{\rm out}} \right),
\end{equation}
where we used the relation,
$e^{\lambda y^-_s}=\frac{3\sqrt{2}\beta\lambda}{\mu^3 \alpha_{\rm in}}
(e^{\frac{4M}{\kappa\lambda}} -1)  $.
Because $\frac{\alpha_{\rm in}}{\alpha_{\rm out}}\gg 1$, the black hole mass
is
\begin{equation}
M \approx \frac{\kappa\lambda}{4}\ln( 2 +\frac{2\mu^6}{9\beta^2\lambda^2\kappa}
   )
\left( \frac{\alpha_{\rm in}}{\alpha_{\rm out}} \right)
\end{equation}
which satisfies the relation (\ref{mass}).
So we may think that the thunderpop can
be a classical outgoing soliton
while the Hawking radiation is a quantum mechanical one.

The finial point to be mentioned is about
the conserved topological charge which
characterizes the soliton.
We apply the method
developed in Ref. [16]
to the topological current
in order to obtain the conservation relation.
The topological soliton charge is defined by
\begin{eqnarray}
\label{qadm}
Q_{\rm Top}(t)&=&\int^{+\infty}_{-\infty}dy^1 J^0(t,y^1) \nonumber \\
              &&=\sqrt{\frac{\beta}{4\mu^2}}
               \left[ f(t,\infty) -f(t,-\infty) \right]
\end{eqnarray}
where the topological current is defined by
$J^{\mu}=\sqrt{\frac{\beta}{4\mu^2}}\epsilon^{\mu\nu}
\partial_{\nu}f$ and we assume that the quantity is defined
in the asymptotically static coordinate for convenience.
In the future null infinity, we define the following Bondi charge
which corresponds to the Bondi energy,
\begin{eqnarray}
\label{qbondi}
Q_{\rm B}(y^-) &=&\frac{1}{2} \int^{+\infty}_{-\infty}  dy^+
                   J^-  (y^+,y^-) \nonumber \\
               &&=\sqrt{\frac{\beta}{4\mu^2}}
               \left[ f(\infty,y^-) -f(-\infty,y^-) \right],
\end{eqnarray}
by integrating the current along the null line.
Then we define the radiation charge which corresponds to the
integrated Hawking radiation,
\begin{eqnarray}
\label{qrad}
Q_{\rm R}(y^-)&=& \frac{1}{2} \int^{y^-}_{-\infty} dy^- J^+
                  (\infty, y^-)\nonumber \\
              &&= \sqrt{\frac{\beta}{4\mu^2}}
              \left[ f(\infty, -\infty) -f(\infty,   y^-) \right].
\end{eqnarray}
By using the eqs (\ref{qadm})-(\ref{qrad}),
the following
conservation relation as an identity holds,
\begin{eqnarray}
Q_{\rm Top}(t) &=& Q_{\rm B}(y^-) +Q_{\rm R}(y^-) +\sqrt{\frac{\beta}{4\mu^2}}
\left[ f(-\infty, y^-) -f(-\infty, \infty)\right]\nonumber \\
&=&Q_{\rm B}(y^-) +Q_{\rm R}(y^-),
\end{eqnarray}
where we used the fact that
our soliton solution (\ref{fsol}) satisfies the condition,
$f(-\infty, y^-)=f(-\infty, \infty)$.

In our model, the soliton charge is 1, which is conserved if
the classical thunderpop is identified with the soliton.
For $y^-<y^-_s$, $Q_{\rm B}(y^-)=1$ and $Q_{\rm R}=0$ since
there is no gradient
for the outgoing soliton solution with
a speed of light.
For $y^->y^-_s$, $Q_{\rm B}(y^-)=0$
and $Q_{\rm R}=1$
due to the gradient of outgoing soliton solution.
So the conservation relation of the topological charge
naturally holds with the help of thunderpop.

We studied the back reaction of the metric due to the Hawking radiation
in the black hole induced by an infalling topological soliton.
The Hawking radiation and Bondi energy were calculated along
the null lines and the sum of them is conserved.
The finial state of the black hole is not different from that
of the RST model. However,
the outgoing (classical) thunderpop may carry an soliton charge.
This situation is natural since the source of classical thunderpop
is given only by the classical soliton in our model whereas the quantum
mechanical radiation may come from the Hawking radiation which
does not carry the topological charge. So the infalling topological
charge may be emitted by the outgoing thunderpop.

\section*{ACKNOWLEDGEMENTS}
We were supported in part by the Korea Science and Engineering Foundation
through the Center for Theoretical Physics (1995) and in part by
the Basic Science Research Institute Program, Ministry of Education,
1995 Project No. BSRI-95-2427.

\end{document}